\documentclass[12pt]{article}
\usepackage{putex}
\usepackage{graphicx}
\begin{document}

\date{\today}
\preprint{hep-th/0505189 \\ PUPT-2163}
\title{Phase transitions near black hole horizons}
\institution{PU}{Joseph Henry Laboratories, Princeton University, Princeton, NJ 08544}
\authors{Steven S. Gubser}
\abstract{The Reissner-Nordstrom black hole in four dimensions can be made unstable without violating the dominant energy condition by introducing a real massive scalar with non-renormalizable interactions with the gauge field.  New stable black hole solutions then exist with greater entropy for fixed mass and charge than the Reissner-Nordstrom solution.  In these new solutions, the scalar condenses to a non-zero value near the horizon.  Various generalizations of these hairy black holes are discussed, and an attempt is made to characterize when black hole hair can occur.}

\maketitle

\tableofcontents

\section{Introduction}
\label{INTRODUCTION}

The conjecture famously expressed in Wheeler's aphorism, ``Black holes have no hair,'' is that black holes are uniquely specified by the conserved quantities that they carry: mass, angular momentum, and charge \cite{Wheeler:1971}.  There are a number of rigorous results on black hole uniqueness and the absence of hair in various theories.  Early results \cite{Israel:1967wq,Israel:1967za,Carter:1971,Wald:1971iw,Bekenstein:1972ny,Bekenstein:1971hc,Bekenstein:1972ky} as well as more recent developments have been reviewed in \cite{Bekenstein:1996pn}.  It is generally recognized that there are significant assumptions made in deriving such results.

A (fairly) precise version of the no-hair conjecture is that for a given mass, angular momentum, and charge, there is a unique, regular, stationary, asymptotically flat black hole solution to four-dimensional classical relativity coupled to sensible matter---that is, matter obeying some positive energy condition that insures that flat space is stable.  The Bartnik-McKinnon solutions \cite{Bartnik:1988am} falsify this version of the conjecture, but these solutions are perturbatively unstable \cite{Straumann:1989tf}.\footnote{There are a number of additional variants and generalizations of the Bartnik-McKinnon solutions which are outside the scope of this paper; see \cite{Volkov:1998cc} for an extended review of the topic.}  It is natural then to guess that perturbatively stable black holes have no hair, in the sense of being unique once locally conserved quantities are specified.  But still there are counter-examples, notably skyrmion hair, which carries baryon number \cite{Luckock:1986tr,Droz:1991cx,Bizon:1992gb} and is argued to be perturbatively stable \cite{Heusler:1992av,Maeda:1993ap} despite being topologically trivial.  (See however \cite{March-Russell:2002fn} for a countervailing argument.)  Also, there are general arguments \cite{Krauss:1988zc} that discrete black hole hair can be generated if there is a discrete gauge symmetry in the low-energy effective theory.

There still seems to be interesting ground to probe between the rigorous results supporting the no-hair conjecture and the counter-examples to its perceived spirit.  In particular, it is natural to inquire whether one can better characterize the conditions under which black hole hair is possible.

In this paper, I construct black hole solutions where a phase transition occurs near the horizon.  This provides a rather general class of asymptotically flat hairy black holes in four dimensions, where the hair is the order parameter of the phase transition.  A novel aspect of this type of hair is that the maximum entropy solution for given conserved quantities (mass and charge in the simplest case) often has hair---meaning that the order parameter is non-zero near the horizon but decays exponentially.  If the phase transition breaks a global symmetry $G$ to a subgroup $H$, then the classical solution with maximum entropy can be continuously non-unique: the solutions are labeled by an element of $G/H$.

An apparent prerequisite to constructing solutions with phase transitions near the horizon is some non-renormalizable coupling in the matter lagrangian.  So it is tempting to speculate in general that black hole hair (suitably defined) can occur only for sufficiently small black holes.  I will return to this theme in section~\ref{CONCLUSIONS}.

Aspects of this paper are related to earlier work.  Through the attractor mechanism \cite{Ferrara:1995ih} one constructs four-dimensional supersymmetric black holes with scalar VEV's controlled by gauge charges and an attractor point.  The attractor point is a special value for massless scalar fields.  For given gauge charges, there can be several (or many) possible attractor points.  All are supersymmetric, and any one of them can be made slightly non-extremal without greatly changing the values of the scalars at the horizon.  This line of thinking already shows that perturbatively stable black holes with given conserved quantities need not be unique.

In section~\ref{EXAMPLE} I exhibit a numerical solution for the simplest case, where the lagrangian has a ${\bf Z}_2$ symmetry which is broken near the horizon through a second order phase transition.  Section~\ref{LIMITS} deals with special limits of this example, and section~\ref{PHASE} presents possible phase diagrams for it.  Section~\ref{GENERALIZE} deals with generalizations of the ${\bf Z}_2$ construction.  Section~\ref{CONCLUSIONS} is an effort to answer a general question: Given some matter lagrangian coupled to gravity, in what range of parameters do hairy black holes typically occur?

\section{A bit of black hair on a charged black hole}
\label{EXAMPLE}

Consider the lagrangian
 \eqn{lagrangian}{
  g^{-1/2} {\cal L} &= {R \over 16\pi G_N} - 
   {1 \over 2} (\partial_\mu\phi)^2 - 
   {f(\phi) \over 4} F_{\mu\nu}^2 - V(\phi)  \cr
  V(\phi) &= {1 \over 2} m^2 \phi^2 
   \qquad f(\phi) = {1 \over 1 + \ell^2 \phi^2} \,.
 }
The aim is to construct static solutions with magnetic charge $g$, defined so that $\int_{S^2} F_2 = 4\pi g$.\footnote{Solutions with electric charge also exist, but they don't exhibit scalar hair when $\ell^2>0$.  Having $\ell^2<0$ threatens to give the gauge field a negative kinetic term, which is pathological.  It's notable that by performing electromagnetic duality on $F_{\mu\nu}$, one winds up with a lagrangian which is entirely quadratic in $\phi$ because $f(\phi) \to 1/f(\phi)$ as part of the duality.  Then it is the electrically charged black holes that can have scalar hair.}  Such solutions must have the form
 \eqn{StaticAnsatz}{
  ds^2 &= g_{tt} dt^2 + g_{rr} dr^2 + 
   r^2 (d\theta^2 + \sin^2\theta \, d\varphi^2)  \cr
  F_2 &= {1 \over 2} F_{\mu\nu} dx^\mu \wedge dx^\nu = 
   g \vol_{S^2} = g \, d\theta \wedge 
   \sin\theta \, d\varphi  \cr
  \phi &= \phi(r) \,.
 }
The relevant equations of motion are
 \eqn{CovariantEOMS}{
  \square\phi &= {1 \over \sqrt{g}} \partial_r \sqrt{g} g^{rr} 
    \partial_r \phi = {\partial V_{\rm eff} \over \partial\phi}  \cr
  V_{\rm eff}(\phi,r) &\equiv V(\phi) + {1 \over 4} f(\phi) 
    F_{\mu\nu}^2 
   = V(\phi) + {g^2 \over 2r^4} f(\phi)  \cr
  G_{\mu\nu} &= 8\pi G_N T_{\mu\nu} \,.
 }
The equations of motion and Bianchi identities for the gauge field are trivially solved by the ansatz $F_2 = g \vol_{S^2}$.  The stress tensor is
 \eqn{Tmunu}{
  T_{\mu\nu} = f(\phi) \left( F_{\mu\lambda} F_\nu{}^\lambda - 
   {1 \over 4} g_{\mu\nu} F_{\alpha\beta}^2 \right) +
   \partial_\mu\phi \partial_\nu\phi - 
   g_{\mu\nu} \left( {1 \over 2} (\partial_\mu \phi)^2 + V(\phi) \right) \,.
 }
Note that the dominant energy condition is obeyed.\fixit{Check against Wald}  With the static magnetic field and scalar profile $\phi(r)$ that enter \CovariantEOMS, one finds
 \eqn{Tcomponents}{
  T^t{}_t &= -{g^{rr} \over 2} \phi'^2 - V_{\rm eff}(\phi,r)  \cr
  T^r{}_r &= {g^{rr} \over 2} \phi'^2 - V_{\rm eff}(\phi,r)  \cr
  T^\theta{}_\theta &= {T^\phi{}_\phi} = 
   {g^{rr} \over 2} \phi'^2 + V(\phi) - {g^2 \over 2r^4} f(\phi)
 }
where primes denote $d/dr$.

It is convenient to set $M_{\rm Pl} \equiv 1/\sqrt{8\pi G_N} = 1$ and to express
 \eqn{gttIs}{
  g_{tt} = -e^{2A(r)} \qquad 
  g_{rr} = e^{2B(r)} \,.
 }

The equations of motion \CovariantEOMS\ lead to four algebraically independent differential equations: the $tt$, $rr$, and $\theta\theta$ Einstein equations, plus the equation of motion for $\phi$.  There is a gauge freedom to rescale $t$ and at the same time shift $A$ by a constant.  So $A$ cannot appear in the equations of motion: only its derivatives can.  A combination of the $tt$ and $rr$ Einstein equations allows one to solve algebraically for $A'$:
 \eqn{LoseA}{\seqalign{\span\TC}{
  (G^t{}_t - T^t{}_t) - (G^r{}_r - T^r{}_r) = 
   -{2 \over r} (A'+B') + g^{rr} \phi'^2 = 0  \cr
  A' = -B' + {r \over 2} \phi'^2 \,.
 }}
After eliminating $A'$ and $A''$ in favor of derivatives of $B$ and $\phi$, three equations of motion remain.  One is a zero energy condition involving only first derivatives of $B$ and $\phi$.  Taking its derivative with respect to $r$ leads to an algebraic combination of the other two equations of motion, one of which can therefore be discarded as redundant.  The final result of these manipulations is the following system of equations:
 \eqn{BphiEOMS}{\seqalign{\span\TC}{
  \phi'' + \left( {2 \over r} - 2 B' + {1 \over 2} r \phi'^2 \right)
   \phi' = e^{2B} 
    {\partial V_{\rm eff} \over \partial\phi}(\phi,r)  \cr
  {1 \over 2} \phi'^2 + e^{2B} V_{\rm eff}(\phi,r) - 
   {2 B' \over r} + {1 - e^{2B} \over r^2} = 0 \,.
 }}
Note that the magnetic charge $g$ enters only through $V_{\rm eff}(\phi,r)$.

The horizon boundary conditions are obtained most simply by considering the Euclidean continuation.  The topology in Euclidean signature is a disk (parametrized by $t$ and $r$) times a sphere (parametrized by $\theta$ and $\phi$), with $F_2$ threading the sphere.  The horizon is the center of the disk, and the boundary conditions are simply that the equations of motion \CovariantEOMS\ are satisfied there.  A series solution near the horizon may be developed:
 \eqn{SeriesH}{
  B &= -{1 \over 2} \log\left( 1 - {r_H \over r} \right) + b_0 + 
     b_1 (r-r_H) + b_2 (r-r_H)^2 + \ldots  \cr
  \phi &= p_0 + p_1 (r-r_H) + p_2 (r-r_H)^2 + \ldots \,.
 }
The series coefficients can be determined recursively in terms of $p_0$:
 \eqn{Coefficients}{
  b_0 &= -{1 \over 2} 
    \log\left( 1 - r_H^2 V_{\rm eff}(p_0,r_H) \right)  \cr
  b_1 &= {4 r_H^4 V_{\rm eff}(p_0,r_H) - 
    4 r_H^6 V_{\rm eff}(p_0,r_H)^2 + 
    3 r_H^6 {\partial V_{\rm eff} \over \partial\phi}(p_0,r_H) 
     \over 8 r_H^3 
     (1 - r_H^2 V_{\rm eff}(p_0,r_H))^2} - 
     {g^2 f(p_0) / 2 r_H^3 \over 1 - r_H^2 V_{\rm eff}(p_0,r_H)}  \cr
  p_1 &= {r_H {\partial V_{\rm eff} \over \partial\phi}(p_0,r_H) 
   \over 1 - r_H^2 V_{\rm eff}(p_0,r_H)} \,.
 }
It may seem remarkable that there is not an additional integration constant for $\phi$, considering that its equation of motion is second order.  The reason (loosely speaking) is that the ``other solution'' has $\phi$ diverging at the horizon.\fixit{Better check this}

With $f(\phi)$, $V(\phi)$, $g$, $r_H$, and $p_0$ specified, one can use the series solutions \SeriesH\ for small $r-r_H$ together with numerics at larger $r$ to integrate the differential equation \BphiEOMS\ up to a chosen upper limit $r_{\rm max}$.  Then the procedure is to ``shoot'' different values of $p_0$ to find the special values where $\phi(r_{\rm max})=0$: for large $r_{\rm max}$, such values are very close to the values that give static solutions which are asymptotic to flat space.  Once a numerical solution is found, the mass of the black hole can be determined by fitting $e^{-2B}$ to the Reissner-Nordstrom form for large $r$---provided $m^2>0$.  The reason for this is that $\phi \to 0$ exponentially fast as $r \to\infty$.  Thus for large $r$ one finds
 \eqn{AsymptoticB}{\seqalign{\span\TC}{
  e^{2B} {g^2 \over 2r^4} + {1-e^{2B} \over r^2} - 
   {2 B' \over r} = 0  \cr
  e^{-2B} = 1 - {M \over 4\pi r} + {g^2 \over 2r^2}
 }}
with corrections that fall off exponentially as $r \to\infty$.

If $m^2=0$ then one must inquire about the higher order behavior of $V(\phi$).  For instance, one could have $V(\phi) = {\lambda \over 4!} \phi^4$ with $\lambda \geq 0$.  Then $\phi$ falls off as a power of $r$ for large $r$, and it becomes slightly trickier to compute the mass.  Here's how to do it.\footnote{Alternatively, and irrespective of the value of $m$, one may use the methods of \cite{BrownYork} to compute the mass of the black hole (see also \cite{BLY} for subsequent developments and a review of relevant literature).}  The ADM mass $M$ is found by comparing the large $r$ asymptotics of $g_{tt}$ to the Schwarzschild solution: in the gauge where $A \to 0$ as $r \to\infty$, we have
 \eqn{FindMass}{
  e^{2A(r)} \sim 1 - {M \over 4\pi r} + O(1/r^2) \quad\hbox{as $r\to\infty$} \,.
 }
So the procedure is to use \LoseA\ together with a known solution $B(r)$, $\phi(r)$ to compute $A(r)$, using the gauge freedom to force $A \to 0$ as $r \to\infty$, and then fit to the form \FindMass.  It should still be true that $B(r) \to 0$ as $r \to\infty$: this is a gauge-invariant statement because it says that space is asymptotically flat rather than asymptotically locally flat with a conical deficit.

Before examining one particular hairy solution in detail, it is instructive to consider a rescaling property of the lagrangian \lagrangian: ${\cal L} \to \Omega^2 {\cal L}$ under the rigid rescaling
 \eqn{RigidConformal}{\seqalign{\span\TL & \span\TR\qquad & 
  \span\TL & \span\TR\qquad & \span\TL & \span\TR}{
  g_{\mu\nu} &\to \Omega^2 g_{\mu\nu} &
   \phi &\to \phi & F_{\mu\nu} &\to \Omega F_{\mu\nu}  \cr
  m^2 &\to \Omega^{-2} m^2 & 
    \ell &\to \ell & g &\to \Omega g
 }}
for constant $\Omega$.  For $\Omega\neq 1$, \RigidConformal\ takes us away from the form of the metric specified in \StaticAnsatz.  It is more convenient to accomplish the rigid rescaling of the metric by sending
 \eqn{KeepMetric}{
  g_{\mu\nu} \to g_{\mu\nu} \qquad r \to \Omega r \qquad
   t \to \Omega t \,.
 }
This is equivalent (by a change of coordinates) to $g_{\mu\nu} \to \Omega^2 g_{\mu\nu}$ with $t$ and $r$ fixed.

Expressed in this latter form, the rigid rescaling clearly has no effect on the equations of motion \BphiEOMS: an overall factor of $\Omega^{-2}$ may be removed from each equation.
But examining \AsymptoticB\ shows that $M \to \Omega M$.  Therefore, given a solution to \BphiEOMS\ with one set of values for $(m^2,\ell^2,g,r_H,M)$, the rigid rescaling produces new solutions with new values $(\Omega^{-2} m^2,\ell^2,\Omega g_0,\Omega r_H,\Omega M)$.  This scaling freedom can be used to set $r_H=1$.  This may seem like an objectionable choice: the horizon is only a Planck length across!  But rescaling with $\Omega$ large leads to solutions with large mass and charge and curvatures that are uniformly small.

\begin{figure}
  \begin{center}
   \includegraphics[width=3in]{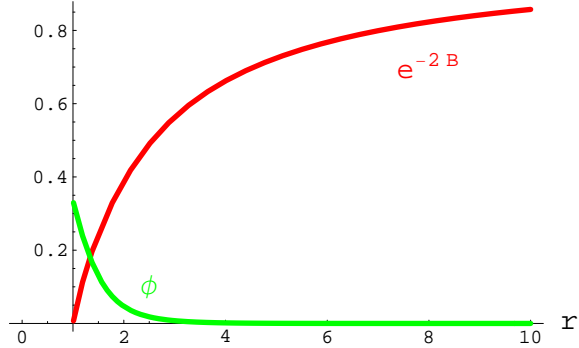}
  \end{center}
  \caption{$g^{rr}$ and $\phi$ for a hairy charged black hole with $m^2=1$, $\lambda=0$, $\ell^2=10$, and $g=1$.}\label{figA}
 \end{figure}

Setting $m^2=1$, $\lambda=0$, $\ell^2 = 10$, $g=1$, and (as planned) $r_H=1$, one quickly arrives at a solution with $\phi \approx 0.3354$ on the horizon.  We exhibit this solution in figure~\ref{figA}.  It has mass $M/4\pi \approx 1.477$, which is lower than $M/4\pi = 1.5$ for the non-extremal Reissner-Nordstrom solution with the same entropy and charge.

\section{Special limits and the phase diagram}
\label{LIMITS}

The obvious question now is for what range of parameters the scalar hair arises.  The lagrangian \lagrangian\ is minimal in the sense that if any of the terms is simplified further, the scalar hair cannot arise.  Having set $M_{\rm Pl}=1$, both of the parameters, $m^2$ and $\ell^2$, are pure numbers.  And having used the scaling symmetry \RigidConformal-\KeepMetric\ to set $r_H=1$, there is no scaling freedom left to adjust $m$ or $g$.  It is easy to see that no other scaling symmetry preserves the equations of motion \BphiEOMS.  The upshot is that the two parameters $m$ and $\ell$ are both meaningful, as is the charge $g$ and the value $p_0$ of the scalar field at the horizon.  We have already seen in section~\ref{EXAMPLE} that demanding an asymptotically flat solution amounts to exactly one constraint.  Thus in the space of possible $(m,\ell,g,p_0)$, there is some co-dimension one hypersurface describing all the static black hole solutions.  This hypersurface has bifurcations, with stable and unstable branches; however, for simplicity I will refer to it as the three-manifold of solutions.  

The aim of this section is to learn about the topology of the three-manifold of solutions.  Determining it through numerics is possible in principle, but CPU-intensive because there are four parameters to scan over.

\subsection{$AdS_2 \times S^2$ solutions}
\label{ADS}

A simple way to motivate searching for $AdS_2 \times S^2$ solutions is to consider the extremal limit of a hairy black hole.  In supersymmetric theories, it has long been understood that a large class of extremal charged black holes are supersymmetric interpolations between flat space and $AdS_2 \times S^2$: this is the topic of the attractor mechanism \cite{Ferrara:1995ih}.  There is no supersymmetry in the lagrangian \lagrangian, but at least it seems reasonable to expect similar interpolating solutions---after all, the standard extremal Reissner-Nordstrom black hole is one.  If there are several $AdS_2 \times S^2$ vacua (labeled by different values of the scalar field), then the expectation is that the corresponding interpolating solutions are hairy extremal black holes, and that the hair survives up to some finite non-extremality.  Extremality amounts to one additional condition on the quantities $(m,\ell,g,p_0)$ that determine static solutions.  So, in the three-manifold of black hole solutions in the space of $(m,\ell,g,p_0)$, the extremal ones should constitute a two-dimensional sub-manifold.

To recap: If an $AdS_2 \times S^2$ vacuum exists, it is expected to be the near-horizon limit of an extremal black hole in asymptotically flat space.  The ansatz for $AdS_2 \times S^2$ vacua is
 \eqn{AdSvacua}{
  ds^2 &= L^2 \left( -\rho^2 dt^2 + 
   {d\rho^2 \over \rho^2} \right) + 
   d\theta^2 + \sin^2\theta \, d\phi^2  \cr
  F_2 &= g \vol_{S^2} \qquad \phi = \hbox{const} \,.
 }
The $S^2$ can be assigned unit radius, as before, by performing a rigid rescaling of a more general ansatz where the $S^2$ has radius $r_H$.  And as before, while the solutions we will describe explicitly have Planck-scale curvatures, rescaling with large $\Omega$ makes the curvatures uniformly small.

The non-trivial equations of motion are
 \eqn{AdSeoms}{
  0 &= {\partial V_{\rm eff} \over \partial\phi}(\phi,1) 
   \equiv {\partial V \over \partial\phi} + 
    {g^2 \over 2} {\partial f \over \partial\phi}  \cr
  -G^t{}_t &= 1 = V_{\rm eff}(\phi,1) \equiv 
   V(\phi) + {g^2 \over 2} f(\phi)  \cr
  G^\theta{}_\theta &= {1 \over L^2} = 
    {g^2 \over 2} f(\phi) - V(\phi) = 1 - 2V(\phi)
 }
where in the last equality, the $G^t{}_t$ Einstein equation has been used.  Treating $m$ and $\ell$ as fixed quantities, the equations \AdSeoms\ have six solutions for $g$ and $\phi=p_0$:
 \eqn{AdSsolns}{\seqalign{\span\TL & \span\TR\qquad & \span\TL & \span\TR}{
  g &= \pm\sqrt{2} & p_0 &= 0  \cr
  g &= \pm {\ell \over m} \left( 1 + {m^2 \over 2\ell^2} \right) &
   p_0 &= \pm {1 \over m} \sqrt{1 - {m^2 \over 2 \ell^2}}
 }}
where in the second line, the two sign choices are independent.  Evidently, the solution exists precisely when $p_0$ is real:
 \eqn{RealityCondition}{
  m^2 r_H^2 < 2\ell^2 M_{\rm Pl}^2 = {\ell^2 \over 4\pi G_N}
 }
where factors of $r_H$ and $G_N$ have been appropriately restored.  The general rule for doing so is to make an expression which is dimensionally correct and which is preserved under the rigid rescaling \RigidConformal.  ($M_{\rm Pl}$ is unaffected by the rigid rescaling.)

So far we have only used the first two lines of \AdSeoms.  The third line gives $L=1$ in the case $p_0=0$, corresponding to the standard near-horizon extremal Reissner-Nordstrom black hole.  For the other values of $p_0$, one obtains
 \eqn{GotL}{
  L = \sqrt{2} \ell {M_{\rm Pl} \over m} = 
   \sqrt{1 \over 4\pi G_N} {\ell \over m} \,.
 }
Thus (for the simple choice $V(\phi) = {1 \over 2} m^2\phi^2$), the third line in \AdSeoms\ and the positivity of $1/L^2$ only require that $\ell$ is non-zero.  (It is assumed in \GotL\ that $\ell>0$.)

It is worth noting a point that is well-understood in the attractor mechanism literature: the mass $M$ and entropy $S = \pi r_H^2/G_N$ of extremal solutions is determined once the charge and the value $p_0$ of $\phi$ at the horizon are chosen.  Both $g$ and $p_0$ take on a discrete set of values: $g$ because of Dirac quantization and $p_0$ because the first two lines of \AdSeoms\ amount to two equations in $g$ and $p_0$.  Explicitly, for the $p_0=0$ solutions, the relation $|g|=\sqrt{2}$ generalizes to $|g| = \sqrt{S}/2\pi$, while for $p_0>0$, the appropriate generalization is
 \eqn{gShair}{
  |g| = {\ell M_{\rm Pl}^2 \over m} + {m \over \ell M_{\rm Pl}^2}
   {S \over 16\pi^2} \geq {\sqrt{S} \over 2\pi}
 }
where the last step follows from the arithmetic-geometric mean inequality.  It's clear from the inequality that for a given charge, the maximum entropy extremal solution (for the theory \lagrangian\ under consideration) is the one with $\phi=0$ everywhere.

\subsection{Static perturbations of the non-extremal Reissner-Nordstrom solution}
\label{PERTURB}

Given that hairy black holes exist, it is natural to expect that the space of static solutions includes at least one bifurcation, where hairy solutions join up with the standard non-extremal Reissner-Nordstrom solution.  Near the bifurcation, the scalar field should be uniformly small, so the hairy solutions can be found perturbatively in $\phi$.  Linearizing the equations of motion \BphiEOMS\ around $\phi=0$ and requiring that the horizon is at $r_H=1$ yields
 \eqn{Linearized}{\seqalign{\span\TC}{
  \phi'' + \left( {2 \over r} - 2 B' \right) \phi' = 
    e^{2B} {\partial^2 V_{\rm eff} \over \partial\phi^2}(0,r) \, 
     \phi  \cr
  B = -{1 \over 2} \log\left( 1 - {M \over 4\pi r} + {g^2 \over 2r^2}
   \right) \qquad
  {\partial^2 V_{\rm eff} \over \partial\phi^2}(0,r) = 
    m^2 - {g^2 \ell^2 \over r^4}  \cr
  \phi \propto 1 + {2 (g^2 \ell^2 - m^2) (r-1) \over g^2-2} \qquad
   \hbox{for $r \to 1$}  \cr
  \phi \propto {e^{-mr} \over r} \qquad
   \hbox{for $r \to\infty$}
 }}
where in the last two lines the normalizable boundary conditions at the horizon and at infinity are shown.  The differential equation cannot be solved in terms of standard special functions.  Before investigating it numerically, it is useful to make a detour into parameter counting and the expected structure of the phase diagram.

Clearly, \Linearized\ constitutes a boundary value problem, and for there to be a solution, one parameter needs to be adjusted.  One way to put it is that for fixed $m$ and $\ell$ (and $r_H=1$), at most discretely many values of $g$ admit solutions to \Linearized: these correspond to bifurcation points of the three-manifold of solutions.  Another way to put it is that for fixed $m_0$ and $\ell$ (again with $r_H=1$), there are at most discretely many values of $g_0$ such that setting
 \eqn{NiceScaling}{
  m = m_0/|g_0| \qquad g = g_0
 }
leads to a normalizable solution to \Linearized.  One may assume $g_0>0$ without loss of generality.  This is a more physically interesting scaling, because such solutions correspond, after a rigid rescaling, to the parameters $(m,\ell,g,r_H) = (m_0/\Omega g_0,\ell,\Omega g_0,\Omega)$.  A black hole which slowly radiates its mass (and entropy) away while conserving its magnetic charge, $g=\Omega g_0$, can then be described by slowly increasing $g_0$ and decreasing $\Omega$ with $\Omega g_0$, $\ell$, and $m_0$ held fixed.

The discussion of the previous paragraph generalizes easily to finite $\phi$: the interesting phase diagram for slowly evaporating black holes comes from fixing $m_0$ and $\ell$ (and $r_H=1$), imposing \NiceScaling, and plotting the value $p_0$ of $\phi$ at the horizon of a static black hole against the charge parameter $g_0$.  For very small $g_0$, no hair is expected, so $p_0=0$: this is the unbroken phase.  For some critical value $g_c$ of $g_0$, there is a solution to the linearized boundary value problem \Linearized, signaling a bifurcation point.  For $g_0>g_c$, solutions with scalar hair should exist, and they spontaneously break the ${\bf Z}_2$ symmetry $\phi \to -\phi$.  The Reissner-Nordstrom black hole is unstable toward developing scalar hair for $g_c < g_0 < \sqrt{2}$.  There might be additional bifurcation points for $g_0>g_c$.  A branch of solutions may terminate on an extremal solution of the type discussed in section~\ref{ADS}.  If so, the limiting value of $g_0$ is determined by plugging \NiceScaling\ into \AdSsolns\ and solving for $g_0$:
 \eqn{gZeroHairy}{
  g_0 = {1 \over \sqrt{2}} \left( {\ell \over m_0} - 
   {\ell^2 \over m_0^2} \right)^{-1/2} \geq \sqrt{2}
 }
with equality precisely if $m_0 = 2\ell$.  Combining \gZeroHairy\ with the last expression for $p_0$ in \AdSsolns, it becomes clear that in order to have a branch of solutions terminate on a regular extremal solution, one needs
 \eqn{ellBounds}{
  {m_0 \over 2} \leq \ell \leq m_0 \,.
 }
To obtain the first inequality in \ellBounds, one starts with the condition that $p_0$ is real (namely $m_0^2 \leq 2\ell^2 g_0^2$) and then uses the equality in \gZeroHairy\ to eliminate $g_0$: the result is an inequality that is equivalent to ${m_0 \over 2} \leq \ell$.  To obtain the second inequality in \ellBounds, one simply requires that the expression for $g_0$ in \gZeroHairy\ is real.  If \ellBounds\ is not satisfied, then the only valid $AdS_2 \times S^2$ solution is the one with $\phi=0$, and it's not obvious how hairy black holes behave as they lose more and more mass at fixed charge.

With these extended preliminaries out of the way, it is time to look at the results of numerics on \Linearized.  

\begin{enumerate}
 \item First consider the case $m_0^2=1$, $\ell^2=10$: above the range specified in \ellBounds.  Values of $g_0$ where a normalizable solution to \Linearized\ was found are:
 \eqn{SomeGValues}{
  g_0/\sqrt{2} &\in \{0.516219,\ 0.714204,\ 0.84679,\ 0.921712,\ 
     0.960911,\ 0.980698,  \cr &\qquad\quad \ 0.990519,\ 0.995355,\ 0.997726\} \,.
 }
The values shown appear to be part of a infinite sequence converging exponentially fast to $g_0=\sqrt{2}$.  The wave-function $\phi(r)$ corresponding to the first entry in \SomeGValues\ has no nodes; the second one has one node; the third has two; and so on.

 \item Now consider the case $m_0^2=1$, $\ell^2=0.9$: within the range \ellBounds.  Values of $g_0$ where a normalizable solution to \Linearized\ was found are:
 \eqn{TwoGValues}{
  g_0/\sqrt{2} \in \{ 0.947941,\ 0.997492 \} \,.
 }
The wave-function $\phi(r)$ corresponding to the first entry in \TwoGValues\ has no nodes, and the second one has one node.  No other values of $g_0$ were found.  However, the numerical problem as stated in \Linearized\ becomes difficult for $g_0 \to \sqrt{2}$ because $d\log\phi/dr$ becomes large near the horizon.  For the points quoted in \TwoGValues, it was checked that the numerics are under control near the horizon; but determining whether or not other solutions exist is hard with the numerical methods employed thus far.

 \item Next consider the case $m_0^2=1$, $\ell^2=0.6$: still within the range \ellBounds.  Only one value of $g_0$ was found where \Linearized\ has a normalizable solution, namely
 \eqn{OneGValue}{
  g_0/\sqrt{2} = 0.989864 \,.
 }

 \item Finally, consider the case $m_0^2=1$, $\ell^2 = 0.1$: below the range specified in \ellBounds.  No values of $g_0$ were found that lead to a normalizable solution for \Linearized.  It is plausible that no hairy solutions with $r_H=1$ exist at all for this choice of $m_0$ and $\ell$.

\end{enumerate}

A familiar tool that hasn't yet been brought to bear is the Breitenlohner-Freedman (BF) bound, which says that a scalar field $\phi$ in $AdS_{d+1}$ of radius $L$ should have a mass satisfying $m_\phi^2 L^2 > -d^2/4$ \cite{Breitenlohner:1982bm,Breitenlohner:1982jf}.  This is the mass entering into the equation of motion: $(\square-m_\phi^2) \phi = 0$.  It's clear from \Linearized\ that for an extremal Reissner-Nordstrom solution, the relevant mass is $m_\phi^2 = {\partial^2 V_{\rm eff} \over \partial\phi^2} (0,1) = m^2 - 2\ell^2$, where we have required $r_H=1$ and used the extremal value $g=\sqrt{2}$.  Two equivalent forms of the BF bound are then
 \eqn{BFbound}{
  m^2 - 2\ell^2 \geq -{1 \over 4}\qquad
  \ell \leq {1 \over 2} \sqrt{m_0^2 + {1 \over 2}} \,.
 }
Cases 1 and 2 above violate the bound \BFbound; cases 3 and 4 do not.

In an asymptotically $AdS_{d+1}$ background, a violation of the BF bound implies that turning on the scalar field can lower the energy.  If the same is true even after the asymptotically $AdS_2 \times S^2$ region is replaced by a transition to flat space, then scalar hair should develop.  The analysis of this paper focuses on static configurations, so instead of a scalar mode with $\omega^2 < 0$, the expected symptom of a BF-violating $AdS_2$ region is an oscillatory $\phi(r)$.  Indeed, in pure $AdS_2$, coordinatized as in \AdSvacua, the oscillatory solutions are $\phi(\rho) = \rho^{-\Delta_\pm}$ where $\Delta_\pm$ are the solutions of $\Delta(\Delta-1)=m_\phi^2 L^2$.  So $\Delta_\pm$ are complex.  Turning on a very slight non-extremality and joining the $AdS_2 \times S^2$ region onto flat space corresponds to imposing a cutoff at small and large $\rho$, respectively.  The closer to extremality one goes, the more wavelengths of $\phi$ oscillations fit between the cutoffs.  So the expected behavior is an infinite discrete series of values of $g_0$ that lead to normalizable solutions of \Linearized.  This should be what is happening in case 2: I expect that only numerical difficulties got in the way of finding more solutions.

If the BF bound \BFbound\ is satisfied in the $AdS_2 \times S^2$ region of the extremal Reissner-Nordstrom solution, it doesn't mean that scalar hair is impossible: it only means that there shouldn't be an infinite discrete series of static near-extremal solutions with oscillations of $\phi$ concentrated near the horizon.  In pure $AdS$, the behaviors of $\phi$ are $\phi(\rho) = \rho^{-\Delta_\pm}$, still with $\Delta_\pm$ solving $\Delta(\Delta-1)=m_\phi^2 L^2$---only this time, $\Delta_\pm$ are real.  A generic combination of these two solutions could have no zeroes or one zero.  So it is reasonable to expect no more than two static perturbations of the Reissner-Nordstrom black hole if \BFbound\ is satisfied.

\section{Suggested phase diagrams}
\label{PHASE}

\begin{figure}
  \centerline{%
   \includegraphics[width=2.5in]{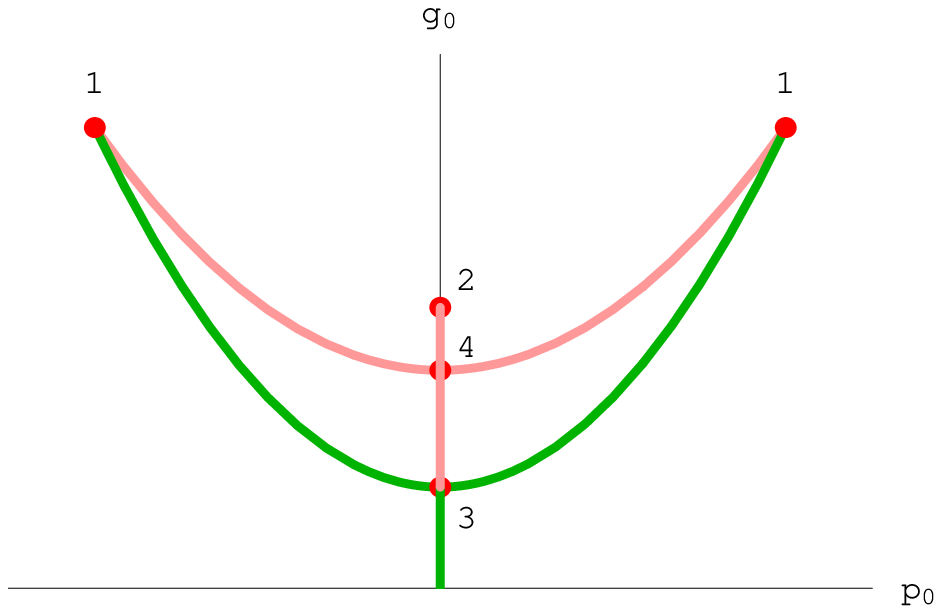}
   \includegraphics[width=2.5in]{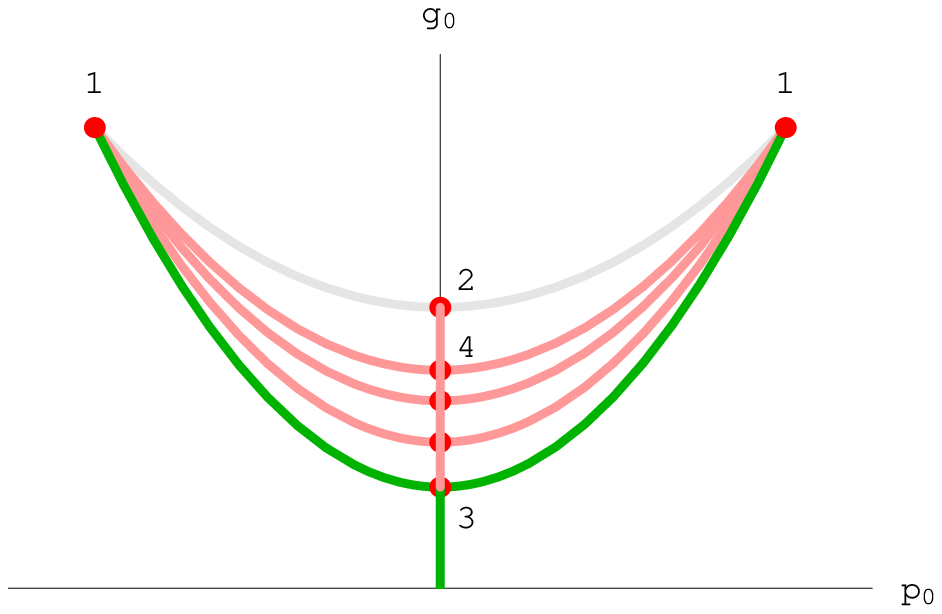}
  }
  \hskip1.7in {\large (A)} \hskip2.3in {\large (B)} \\[10pt]
  \centerline{%
   \includegraphics[width=2.5in]{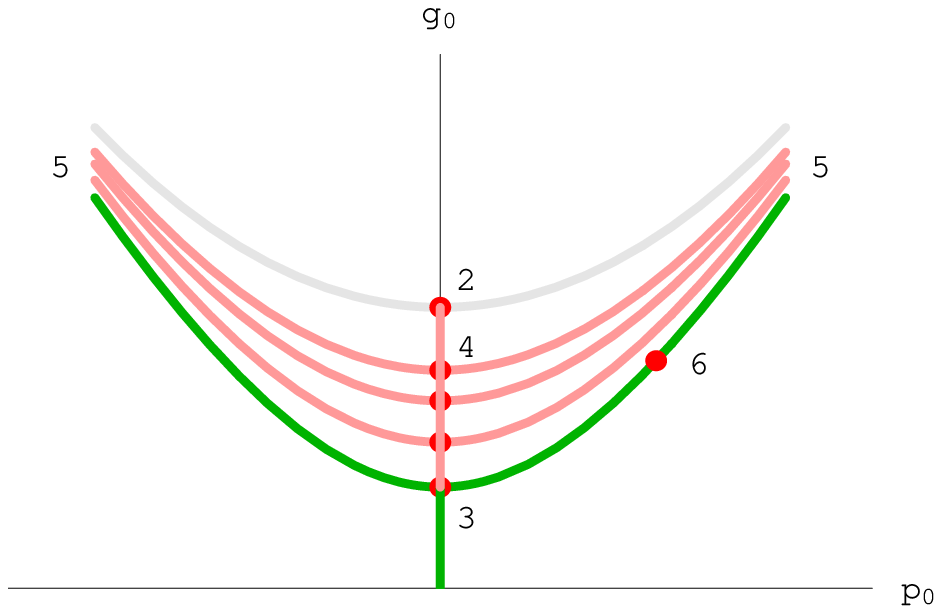}
  }
  \centerline{\large (C)\hskip0.25in}
  \caption{(A) The suggested phase diagram for ${m_0 \over 2} \leq \ell < {1 \over 2} \sqrt{m_0^2 + {1 \over 2}}$.  (B) The suggested phase diagram for ${1 \over 2} \sqrt{m_0^2 + {1 \over 2}} < \ell < m_0$.  (C) The suggested phase diagram for $m_0 < \ell$.  The green lines indicate stable branches, and the oranges lines indicate unstable branches.  The gray line in (B) and (C) is the asymptotic behavior of many unstable branches.  Properties of the numbered points are discussed in the main text.}\label{figB}
 \end{figure}

All the evidence so far is consistent with the following picture of the phase diagram (taking $p_0>0$ and $g>0$ without loss of generality):
 \begin{itemize}
  \item Hairy black holes occur when $\ell M_{\rm Pl} > {mg \over 2M_{\rm Pl}}$.  This is the lower bound in \ellBounds.
  \item There is only one branch of stable hairy black holes, and on this branch, $\phi(r)$ has no nodes.  Considering $\ell$ and $m$ as fixed parameters, black holes on this branch have the largest possible $p_0$ for given black hole mass and charge.  When such black holes exist, the Reissner-Nordstrom solution with the same charge and entropy is unstable.
  \item If $\ell M_{\rm Pl} > {1 \over 2} \sqrt{\left( {mg \over M_{\rm Pl}} \right)^2 + {1 \over 2}}$ (which comes from violating the BF bound \BFbound) then there are infinitely many branches of (unstable) static black hole solutions splitting off from the Reissner-Nordstrom solution.  Otherwise, there are at most two.
  \item If ${mg \over 2M_{\rm Pl}^2} < \ell M_{\rm Pl} \leq {mg \over M_{\rm Pl}}$ (a translation of \ellBounds) then all branches of hairy black hole solutions end on the hairy extremal solution described in section~\ref{ADS}.  Otherwise, they terminate on a naked singularity.
 \end{itemize}
Figure~\ref{figB} shows my guess of what the phase diagram looks like in $p_0$-$g_0$ planes with various fixed values of $m_0$ and $\ell$.
Various points on these suggested phase diagrams deserve special mention:
 \begin{enumerate}
  \item The near-horizon geometries at these points are the $AdS_2 \times S^2$ solutions with $p_0 \neq 0$ described in \AdSsolns.
  \item These points represent the standard extremal Reissner-Nordstrom solutions.  The lines extending vertically down from them represent the non-extremal Reissner-Nordstrom solutions.
  \item These points correspond to the first solutions of the perturbative equation \Linearized\ for $\phi$---the solutions with no nodes.
  \item These points correspond to other solutions of the perturbative equation for $\phi$.  Many solutions were found for the case shown in figure~\ref{figB}(C).  Only two solutions were found for the case shown in figure~\ref{figB}(B), but more are expected based on the argument following \BFbound.  For the case shown in figure~\ref{figB}(A), depending on the precise value of $m_0$ and $\ell$, there might be only one solution.  If so, the orange curve passing through point 4 should be deleted.
  \item There are no $AdS_2 \times S^2$ solutions for the case shown in figure~\ref{figB}(C): see \gZeroHairy\ and \ellBounds.  I do not know the precise asymptotic behavior of the stable and unstable branches.
  \item This point represents the black hole solution exhibited in figure~\ref{figA}.
 \end{enumerate}

\section{Generalizations of the ${\bf Z}_2$ case}
\label{GENERALIZE}

One can envision a large class of generalizations of the basic construction explained in section~\ref{EXAMPLE}.  If a certain lagrangian exhibits a phase transition as one tunes its parameters, then by adding appropriate couplings to $F_{\mu\nu}^2$ of a $U(1)$ gauge field, one can arrange to pass through the phase transition near the horizon of a charged black hole.  What drives the phase transition is the intensity of the electromagnetic fields, which are used as a control parameter for the phase transition.  If an analog of the rigid rescaling \RigidConformal\ can be defined, then curvatures and field strengths can be made uniformly small outside the horizon by suitably scaling parameters in the lagrangian.  The horizon is not a special or singular location in this type of classical construction.  The phase transition occurs in a finite region close to the horizon.

An obvious direction in which to generalize the basic ${\bf Z}_2$ example is to have several $U(1)$ gauge fields and several scalars:
 \eqn{SeveralFields}{
  g^{-1/2} {\cal L} &= {R \over 16\pi G_N} - 
   {1 \over 2} (\partial_\mu \vec\phi)^2 - 
   \sum_a {f_a(\vec\phi) \over 4} (F^a_{\mu\nu})^2 - V(\vec\phi)  \cr
  V_{\rm eff}(\vec\phi,r) &= V(\vec\phi) + \sum_a
   {g_a^2 \over 2r^4} f_a(\vec\phi)
 }
where $g_a$ are the magnetic charges under the gauge fields $F^a_{\mu\nu}$ of a spherically symmetric black hole.  Imitating the logic of section~\ref{ADS}, possible extremal hairy solutions correspond to $AdS_2 \times S^2$ vacua of the theory \SeveralFields, and these correspond to extrema of $V_{\rm eff}(\vec\phi,r_H)$ where $r_H$ is the horizon radius.  \footnote{These extrema are attractor points in the context of ${\cal N}=2$ compactifications of string theory to four dimensions.  In that case, $V(\vec\phi)=0$, so the attractor points do not depend on $r_H$.}  If there are $N$ scalars with similar masses and couplings, and if one can expect some $2^N$ minima of $V_{\rm eff}(\vec\phi,r)$ within some finite range of the origin, then it is sensible to think that the number of black holes with the same mass and charge is also exponential in $N$.  This of course is a similar counting to landscape estimates of the number of flux vacua \cite{Bousso:2000xa}.

In the rest of this section, several other generalizations of the basic ${\bf Z}_2$-breaking example will be discussed.

\subsection{Finite Hawking temperature}
\label{HAWKING}

It is interesting to inquire how the hairy black hole construction of section~\ref{EXAMPLE} interacts with the finite Hawking temperature of a black hole horizon.  The effective temperature $T_{\rm local} = \sqrt{|g^{tt}|} T_H$ diverges at the horizon, so one might expect a restoration of symmetry due to high-temperature effects sufficiently close to the horizon.  The Hawking temperature is a quantum effect, so if the horizon is large compared to the Planck scale, finite temperature effects are significant only in some region where $(r-r_H)/r_H \ll 1$.  Thus the hairy solutions described in sections~\ref{EXAMPLE}, \ref{LIMITS}, and~\ref{PHASE} seem likely to survive quantum corrections.\footnote{For hairy black holes with an extremal limit involving an $AdS_2 \times S^2$ near-horizon geometry, the Hawking temperature vanishes at extremality: then the considerations of this section do not apply.}

Let's explore the effects of finite Hawking temperature more quantitatively by setting
 \eqn{NewVeff}{
  V_{\rm eff}(\phi,r) = {1 \over 2} m^2 \phi^2 + 
   {1 \over 4} {F_{\mu\nu}^2 \over 1+\ell^2\phi^2} + 
   {1 \over 2} \lambda T_{\rm local}^2 \phi^2
 }
in \BphiEOMS.  Here $\lambda$ is proportional to a small positive quartic coupling for $\phi$ which is assumed to be otherwise unimportant in \NewVeff.  The presence of $T_{\rm local}$ introduces dependence on $A(r) = {1 \over 2} \log |g_{tt}|$, so one must supplement \BphiEOMS\ by the second equation in \LoseA\ to obtain a well-posed set of three differential equations for $A$, $B$, and $\phi$.  Let's consider the situation where the $\lambda T_{\rm local}^2$ term is small except very close to the horizon, and let's also restrict to small $\phi$, so that we may use the linearized equation for $\phi$:
 \eqn{LinLTeom}{
  \phi'' + \left( {2 \over r} - 2 B' \right) \phi' = 
    e^{2B} \left[ 
     m^2 - {\ell^2 \over 2} F_{\mu\nu}^2 + 
      \tilde\lambda e^{-2A} \right] \phi
 }
where $\tilde\lambda = \lambda T_H^2$.  The aim is to see that for given $m^2$ and $\ell^2$, and for $r_H=1$, the values of $g$ where \LinLTeom\ has a normalizable solution carry over smoothly to the results presented in section~\ref{PERTURB} as $\tilde\lambda \to 0$.  Analyzing the near-horizon asymptotics of \LinLTeom\ leads to
 \eqn{phiRh}{
  \phi(r) \sim (r-1)^{2\sqrt{\tilde\lambda} / (2-g^2)}
 }
for $r-1 \ll 1$.  Recall that $|g| \leq \sqrt{2}$ in order to have a regular horizon.  When $\tilde\lambda \ll 1$ (and for fixed $g > \sqrt{2}$) the power of $r-1$ in \phiRh\ is very small, so \phiRh\ is almost like setting $\phi$ at the horizon to a constant.  Indeed, the results of numerics show that the values of $g$ where \LinLTeom\ admits normalizable solutions are only slightly different when $\tilde\lambda$ is small than when $\tilde\lambda = 0$.  This reinforces the claim that the hairy black hole solutions described in previous sections survive quantum corrections.

\subsection{Larger symmetry groups and gauged symmetries}
\label{GAUGED}

Next let's see how the classical ${\bf Z}_2$-breaking solution can be embedded into theories with larger symmetry groups.  For example, consider an $O(N)$-symmetric theory where the lagrangian is
 \eqn{ONL}{
  g^{-1/2} {\cal L} &= {R \over 16\pi G_N} - 
   {1 \over 2} (\partial_\mu\vec\phi)^2 - 
   {f(|\vec\phi|) \over 4} F_{\mu\nu}^2 - V(|\vec\phi|)  \cr
  V(|\vec\phi|) &= {1 \over 2} m^2 \vec\phi^2 
   \qquad f(|\vec\phi|) = {1 \over 1 + \ell^2 \vec\phi^2} \,.
 }
The ${\bf Z}_2$-breaking solution can be embedded in this $O(N)$-symmetry theory by choosing one component of $\vec\phi$ as the one that becomes non-zero at the horizon.  Hairy black hole solutions now preserve $H = O(N-1) \subset O(N) = G$, and the moduli space of solutions is $G/H = O(N)/O(N-1) = S^{N-1}$.

One might naturally expect that there are zero-modes of black holes labeled by an element of a continuous coset space $G/H$.  In the case $N=2$, where $G/H = S^1$, one might think that exciting such modes could lead to black holes with an arbitrary charge under $G \approx U(1)$.  In fact, there are subtleties in this argument, which I will return to in section~\ref{CHARGES}.

So far, the symmetries that get spontaneously broken near a black hole horizon have been global.  But there are good reasons to think that in theories that include gravity, all genuine symmetries are gauged.  This has interesting consequences.  Suppose first that there is no gauging of the $O(N)$ symmetry of \ONL.  Then gravitational effects are supposed to break $O(N)$ explicitly, so $V_{\rm eff}(\vec\phi,r)$ no longer has degenerate minima.  If the explicit symmetry breaking is small, then for parameters such that a phase transition can occur near the horizon, a generic circumstance is for there to be three static solutions.  Only one is stable, namely the one corresponding to the true minimum of $V_{\rm eff}(\vec\phi,r)$.  Another, which is unstable, corresponds to the unbroken phase, where $\vec\phi \approx 0$ everywhere outside the horizon.  A third, also unstable, corresponds to the saddle point of $V_{\rm eff}(\vec\phi,r)$ which is the highest energy point in the coset space of points that were degenerate minima before the explicit symmetry breaking.  But if we return to the ${\bf Z}_2$-symmetric model, then this last type of solution is perturbatively stable.

Now suppose instead that the $O(N)$ symmetry of \ONL\ is gauged.  Then $V_{\rm eff}$ can only be a function of $|\vec\phi|$ and $r$.  The ${\bf Z}_2$-breaking solution can be embedded into the gauged $O(N)$-symmetry theory as before (by distinguishing an element of $G/H = S^{N-1}$) but every direction in the coset space is gauge-equivalent.  Thus there is only one hairy black hole solution, at least when we ignore the possibility of turning on non-zero field strengths of the $O(N)$ gauge fields.

\subsection{First order behavior}
\label{FIRSTORDER}

I have focused on the simplest case of a second order phase transition, but first order behavior is possible too.  To explore this, let us again examine an explicit example.  The lagrangian is just as in \lagrangian, except that
 \eqn{NewF}{
  f(\phi) = {1 \over 1 + \ell_1^2 \phi^2} -
    {\kappa \over 1 + \ell_2^2 \phi^2} \,.
 }
The parameters $\ell_1$, $\ell_2$, and $\kappa$ must be chosen so that $f(\phi) \geq 0$ for all $\phi$ (or at least all $\phi$ which occur in the solution); otherwise the gauge kinetic term can acquire a wrong sign.  As in the previous examples, it is convenient to set $M_{\rm Pl}=1$.  There is a scaling symmetry similar to \RigidConformal, but none of the parameters $\ell_1$, $\ell_2$, and $\kappa$ transform under it.  It is convenient to set $r_H=1$ while recalling that the scaling symmetry may be used to generate solutions with arbitrarily large horizons.  A suitable choice of the other parameters is
 \eqn{SuitableChoice}{
  \ell_1^2 = 3 \,,\quad \ell_2^2 = 25 \,,\quad 
   \kappa = 1/5 \,,\quad g/\sqrt{2} \gsim 0.86 \,.
 }
Only for values of $g$ as described in \SuitableChoice\ were hairy solutions found.  For any given $g$ in this range, there are five solutions: the Reissner-Nordstrom solution, with $\phi=0$ everywhere; two distinct solutions with $\phi>0$ everywhere; and the ${\bf Z}_2$ images of these two hairy solutions.

\begin{figure}
  \centerline{\includegraphics[width=2.6in]{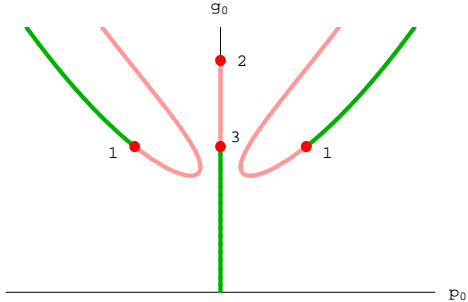}}
  \caption{First order behavior in a phase transition near a black hole horizon.  The parameters $m$, $\ell_i$, and $\kappa$ are held fixed, with values as in \SuitableChoice, and the $p_0$-$g_0$ plane is as described after \NiceScaling.  The green lines from point $3$ to the origin and outward from the points labeled $1$ represent the lowest mass solution for given charge and entropy, while the orange lines represent other static solutions.  The numbered points are explained in the main text.}\label{figC}
 \end{figure}

For $0.86 \lsim g/\sqrt{2} \lsim 0.93$, the hairy solutions are heavier than the Reissner-Nordstrom solution with the same charge and horizon radius.  For $0.93 \lsim g/\sqrt{2} \lsim 1$, one of the hairy solutions is lighter than the Reissner-Nordstrom solution, and the other is heavier.  For example, when $g/\sqrt{2} = 0.97$, the masses of the Reissner-Nordstrom black hole and the two hairy black holes stand in the ratios $1 \, : \, 1.0023 \, : \, 0.9950$.  The situation is summarized qualitatively in figure~\ref{figC}.
The numbered points in figure~\ref{figC} are:
 \begin{enumerate}
  \item This hairy solution is degenerate in mass with the Reissner-Nordstrom solution with the same charge and entropy, labeled $3$.
  \item The family of Reissner-Nordstrom black holes terminates at finite $g_0$ on an extremal black hole.  I have not investigated how the other branches of solutions might terminate.
  \item As explained in $1$.
 \end{enumerate}

In the theory specified by \lagrangian\ and \NewF, the Reissner-Nordstrom black hole is always perturbatively stable.  This is not a necessary feature: if $f(\phi)$ has a small negative second derivative at the origin and $V(\phi)$ is chosen appropriately, then the Reissner-Nordstrom solutions can become unstable at a larger value of $g_0$ than when the first hairy black hole solutions appear.  A natural guess for the phase diagram then is that the inner legs of the hairy branches shown in figure~\ref{figC} join up with the Reissner-Nordstrom solutions.

The ${\bf Z}_2$ symmetry is inessential when one is interested in first order transitions.  In fact, if one starts with the ${\bf Z}_2$-symmetric theory \lagrangian\ and breaks the ${\bf Z}_2$ symmetry slightly (say by adding a small $\phi^3$ term to $V(\phi)$), the resulting solutions will still exhibit some first order phase transition.

\subsection{Can a black hole carry a global Noether charge?}
\label{CHARGES}

We have seen that under suitable conditions, black holes can have hair in the sense of there being a scalar field which condenses near the horizon and spontaneously breaks a symmetry.  If that symmetry is a global $U(1)$, the obvious next question is whether one can excite a collective coordinate of the static black hole to make a black hole that carries a global $U(1)$ Noether charge.

The answer seems to be no.  The arguments in support of this claim will be somewhat heuristic.  See also \cite{March-Russell:2002fn}, where a different line of argument was advanced favoring the same conclusion.

In formulating an ansatz for a black hole that carries a $U(1)$ charge, one would naturally try separation of variables: $\phi(t,r) = e^{-i\omega t} \phi(r)$.  Then the $U(1)$ charge density would be an integral of 
 \eqn{ChargeDensity}{
  J_t = {1 \over 2i} \phi^* \overleftrightarrow\partial_t \phi \,.
 }
That is, $J_t = -\omega |\phi|^2$.  In section~\ref{EXAMPLE}, we saw that the scalar is finite at the horizon in a static solution.  This presumably persists for non-zero $\omega$, because the value of $\phi$ at the horizon is one of the shooting parameters that allows the construction of hairy solutions.  The trouble is that the correct horizon boundary conditions are that $\phi(t,r)$ should be a function only of the infalling coordinate at the horizon.  This is equivalent to saying that there is no outgoing flux of $\phi$ in the classical solution.  There is a net infall of energy of order $\omega^2 r_H^2 |\phi|^2_{\rm horizon}$ where $\omega$ is a typical frequency of the solution.  Two powers of $\omega$ are involved because $T_{\mu\nu}$ involves two derivatives, and $r_H^2$ comes in because the energy flux is integrated over the horizon.  The total energy in a time-variable ansatz can be estimated as $\omega^2 r_H^2 L |\phi|^2_{\rm horizon}$ plus the energy of the static solution.  Here $L$ is the radial extent of the scalar hair.  If this estimate is correct, then the time-dependence decays on a time scale of $L$ itself: the time-dependent part of the fields falls into the black hole at roughly the speed of light.

Clearly, the infalling boundary conditions at a horizon are crucial to the discussion.  In Euclidean signature, the horizon becomes a smooth point---namely the point where $|\phi|$ is maximized in the case of a black hole solution with complex scalar hair.  It seems clear that the Euclidean solution can be generalized to a stationary geometry where the scalar evolves with a factor $e^{i \omega t_E}$, where $\omega$ is arbitrary (at least in some neighborhood of $\omega=0$).  Attempting to Wick rotate such configurations back to Minkowski signature gives exponentially increasing or decreasing scalar profiles which do not change the phase of $\phi$.

For black hole horizons that are extended only in the time direction, a natural expectation is that complex scalar hair does not have infinite-time phase coherence.  That is, low-energy fluctuations probably make $\langle \phi^*(\vec{x},t) \phi(\vec{x},t') \rangle$ decay at least as some inverse power of $|t-t'|$, at least for generic spatial locations $\vec{x}$.  Perhaps the stationary Euclidean solutions with $\phi \sim e^{i \omega t_E}$ are in some sense responsible for this expected Coleman-Mermin-Wagner behavior.

Is it possible to design a black hole solution which can support a global Noether charge?  Perhaps one can at least come closer than the examples discussed above.  If, in the classical solution, $\phi$ approaches $0$ both at infinity and at the horizon but is non-zero in between, then low-energy excitations would be discouraged from falling into the horizon, simply because $\phi_{\rm horizon}$ is still zero (or at least small) for the excited state.  Thus one might hope for quasi-normal modes with long lifetimes that carry a global Noether charge.

In section~\ref{HAWKING} it was argued that finite Hawking temperature causes symmetry to be restored at the horizon: $\phi \to 0$ as some small positive power of $r-r_H$.  But this would {\it not} suppress the infall of time-dependent excitations into the black hole.  The reason is that the rate of infall is proportional to $\langle J_\mu \rangle$, which is a bilinear in $\phi$, and hence more closely related to (a suitable regularization of) $\langle \phi^2 \rangle$ than $\langle \phi \rangle^2$.  And $\langle \phi^2 \rangle$ is not suppressed by the finite temperature effects.  $\langle\phi\rangle \to 0$ near the horizon because the high temperature causes strong fluctuations of the scalar there.  But the same strong fluctuations mean that $\langle \phi^2 \rangle$ is at least comparable to $\phi^2$ in the classical solution that ignores Hawking temperature.\footnote{This line of argument is reminiscent of those in \cite{March-Russell:2002fn}.}  By the same token, if a classical solution has $\phi \to 0$ at the horizon (as proposed in the previous paragraph) but fluctuations of $\phi$ at the horizon are strong because of the local Hawking temperature, then quasi-normal modes will {\it not} be particularly long-lived.

In conclusion, there are substantial obstacles to constructing black holes that carry a global Noether charge, stemming (heuristically) from the tendency of time-dependent excitations of a static black hole configuration to fall through the horizon.  These arguments do not apply to topological currents $j^\mu$ like the one which defines the baryon number of Skyrmions.  In that setup, entirely static field configurations can have $j^0 \neq 0$.

\subsection{Asymptotically anti-de Sitter solutions}
\label{ADSPHASE}

Hairy black holes are commonplace in anti-de Sitter space.  Examples in $AdS_5$ with translation invariance in three spatial directions as well as time were described in \cite{Gubser:2000nd}, where it was also shown that a necessary condition for such solutions to exist in a theory with a two-derivative action is that the scalar potential at the horizon should be no greater than at asymptotic infinity.  This condition might be close to sufficient for black holes with translation invariance to exist: certainly a counting of free parameter suggests this.  An example was given in \cite{Gubser:2000nd} based on dimension two operators in the field theory.  Subsequent developments of hairy anti-de Sitter black holes are summarized in \cite{Sudarsky:2002mk,Torii:2001pg,Winstanley:2002jt,Hertog:2004dr,HertogMaeda}.

The simplest version of a hairy black hole in $AdS_{d+1}$ involves just one real scalar and has the following form:
 \eqn{AdSansatz}{
  S &= \int d^{d+1} x \, \sqrt{g} \left[ R 
   - {1 \over 2} (\partial_\mu \phi)^2 - V(\phi) \right]  \cr
  ds^2 &= e^{2A(r)} (-h(r) dt^2 + d\vec{x}^2) + {dr^2 \over h(r)}
   \qquad \phi = \phi(r)
 }
where $V(\phi)$ has a maximum at $\phi=0$ with $V(0) < 0$ and $V''(r) < 0$.  Thus $\phi$ is a tachyon whose mass determines the dimension of a dual operator ${\cal O}$ in the field theory.  Asymptotically, as $r \to \infty$, one requires $A(r) \to r/L$, $h(r) \to 1$, and $\phi \to e^{(\Delta-d) r/L} + \kappa e^{-\Delta r/L}$ where $\Delta$ is the dimension of ${\cal O}$ and $\kappa$ is fixed by the solution and is proportional to $\langle {\cal O} \rangle$.  The black hole horizon occurs at the largest value of $r$, call it $r_H$, where $h(r)=0$.  Black holes of the form \AdSansatz\ (with non-trivial $\phi(r)$) should exist, at least for large enough $r_H$, because they represent a thermal state of the CFT deformed by the relevant operator ${\cal O}$.  The story is only slightly changed if one replaces $d\vec{x}^2$ by a constant curvature $(d-1)$-manifold: still, for large enough horizon radii, hairy black holes should exist.  The dual description is the ${\cal O}$-deformed CFT on the curved $(d-1)$-manifold at finite temperature.  It is harder to say what happens as $r_H$ is decreased.  In rough terms, it may happen that a singularity in $A$ or $\phi$ is ``uncovered'' when $r_H$ decreases below a certain bound.  In \cite{Gubser:2000nd} it was proposed that all such singularities have to be physical because they correspond to configurations that can be reached through a process of black hole evaporation.

There are two reasons not to regard these hairy solutions as significant exceptions to the no-hair conjectures.  First, boundary conditions are imposed on the scalar at asymptotic infinity which are incompatible with $\phi=0$ everywhere---so the hair is not so much a property of the black hole horizon as of asymptotic infinity.  Second, with those boundary conditions imposed, the black hole solution is typically unique.\footnote{See however \cite{HertogMaeda}, where a non-linear boundary condition corresponding to a triple-trace term is imposed which {\it is} compatible with $\phi=0$ and admits non-unique black hole solutions in $AdS_4$.  This construction is special because the mass of the scalar has to take on a certain special value, corresponding to an operator in the dual field theory of dimension $1$.}  

In the spirit of the other constructions in this paper, it would be interesting to see black hole solutions in $AdS_{d+1}$ which are non-unique below a certain temperature and in which hair develops near the horizon.  An example is provided by the following action in $AdS_5$:
 \eqn{HybridAdS}{\seqalign{\span\TC}{
  S = \int d^5 x \, \sqrt{g} \left[ R 
   - {1 \over 2} (\partial_\mu \phi)^2 
   - {1 \over 2} (\partial\chi)^2 - V(\phi,\chi) \right]  \cr
  V(\phi,\chi) = -{6 \over L^2} + 
   {1 \over 2} m_\phi^2 \phi^2 + {1 \over 2} m_\chi^2 \chi^2 + 
    {g \over 4} \phi^2 \chi^2
 }}
with $-4 < m_\phi^2 L^2 < 0$ and $g < 0$.  One imposes the following boundary conditions for large $r$:
 \eqn{HybridBCs}{
  A \to r/L \qquad
  \phi \to e^{(\Delta_\phi-4) r/L} + 
    \kappa_\phi e^{-\Delta_\phi r/L} \qquad
  \chi \to \kappa_\chi e^{-\Delta_\chi r/L}
 }
where $\Delta_\phi (4-\Delta_\phi) = m_\phi^2 L^2$ and $\Delta_\chi (4-\Delta_\chi) = m_\chi^2 L^2$---the larger root for $\Delta$ being chosen in both cases.  The constants $\kappa_\phi$ and $\kappa_\chi$ are to be determined by the solution.  There is a ${\bf Z}_2 \times {\bf Z}_2$ symmetry of the action \HybridAdS, and the ansatz either breaks to ${\bf Z}_2$ (if $\kappa_\chi = 0$) or to nothing (if $\kappa_\chi \neq 0$).

The field theory has some strongly coupled UV fixed point, dual to pure $AdS_5$.  A renormalization group flow is started by including a relevant deformation $\Lambda_\phi^{4-\Delta_\phi} {\cal O}_\phi$ explicitly in the lagrangian, where $\Lambda_\phi$ is an energy scale---the scale where the relevant deformation becomes important.  Setting the coefficient of $e^{(\Delta_\phi-4) r/L}$ equal to unity in \HybridBCs\ amounts to choosing units so that $\Lambda_\phi = 1$ (or, more precisely, some order unity, $T$-independent constant).  The quantity 
 \eqn{kappaChiIs}{
  \kappa_\chi = {\langle {\cal O}_\chi \rangle \over 
    \Lambda_\phi^{\Delta_\chi}}
 }
is the order parameter of spontaneous symmetry breaking.  (The equality in \kappaChiIs\ ignores another order unity, $T$-independent constant factor, as do subsequent formulas in this section.)  At high temperatures, one expects $\langle {\cal O}_\chi \rangle = 0$, while at low temperatures, one expects $\langle {\cal O}_\chi \rangle \neq 0$.  

Units can be chosen on the supergravity side so that $L=1$; thus it is not a meaningful parameter classically.  There are in fact three parameters in the action: $m_\phi^2$, $m_\chi^2$, and $g$; or, equivalently, $\Delta_\phi$, $\Delta_\chi$, and $g$.  Having chosen these parameters and imposed the boundary conditions \HybridBCs, there is only one more meaningful parameter that may be freely adjusted in black hole solutions.  This parameter can either be the Hawking temperature or the value of $\phi$ at the horizon.  When it too is chosen, then $\chi$ at the horizon as well as $\kappa_\phi = \langle {\cal O}_\phi \rangle / \Lambda_\phi^{\Delta_\phi}$ and $\kappa_\chi = \langle {\cal O}_\chi \rangle / \Lambda_\phi^{\Delta_\chi}$ are determined, possibly up to discrete choices.

\begin{figure}
  \centerline{\includegraphics[width=3in]{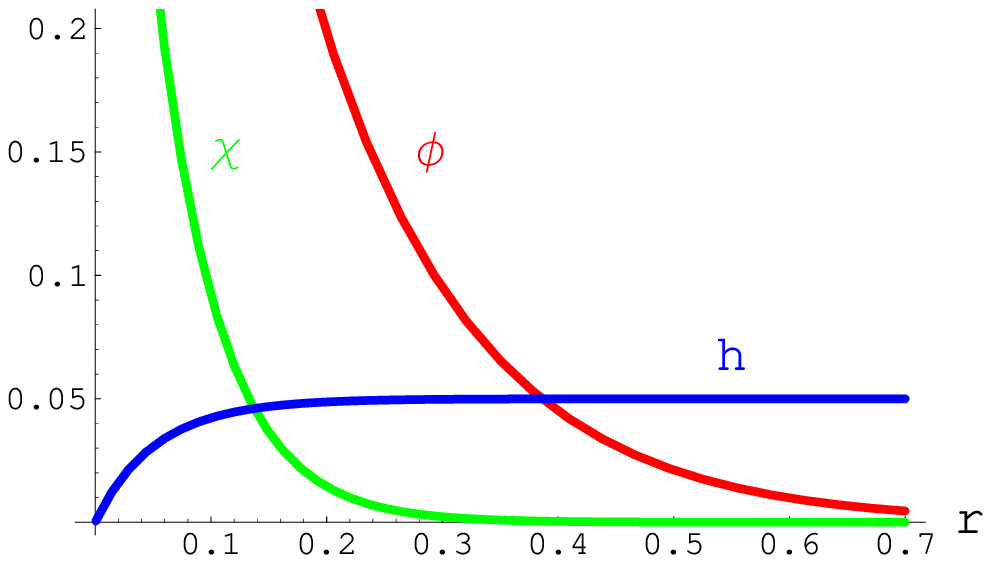}
   \includegraphics[width=3in]{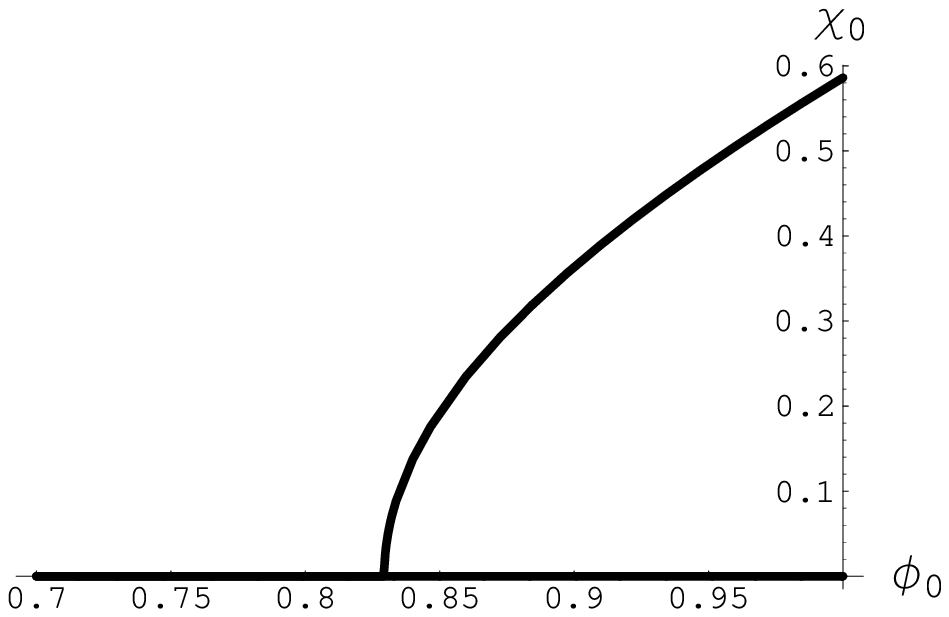}}\ \\
  \centerline{{\large (A)} \hskip2.3in {\large (B)}}
  \caption{A phase transition near a black hole horizon in $AdS_5$.  (A) The functions $\phi(r)$, $\chi(r)$, and $h(r)$ for the hairy solution with $\phi = 1$ at the horizon.  (B) $\phi_0$ and $\chi_0$ are the values of $\phi$ and $\chi$ at the horizon.  Hairy horizons evidently exist only for $\phi_0 > 0.83$.}\label{figD}
 \end{figure}

For a preliminary exploration, let's choose $\Delta_\phi = 2.2$, $\Delta_\chi = 4.2$, and $g = 10$.  Then one may plot the value $\chi_0$ of $\chi$ at the horizon against the value $\phi_0$ of $\phi$ at the horizon and find the branching behavior characteristic of a second order phase transition.  See figure~\ref{figD}.

In constructing the black hole solutions that go into figure~\ref{figD}, it is most straightforward to set the initial conditions at the horizon, which by convention can be taken to be at $r=0$:
 \eqn{HorizonHybrid}{
  h'(0) = 1 \,,\qquad \phi(0) = \phi_0 \,,\qquad \chi(0) = \chi_0 \,.
 }
The procedure then is to use some combination of series solutions and numerics to integrate the equations of motion following from the action \HybridAdS\ to large $r$, and then achieve $\chi \to 0$ by appropriately tuning $\chi_0$.  (If $m_\chi^2 < 0$, then more subtle boundary conditions have to be imposed on $\chi$ at large $r$.)  Once a suitable $\chi_0$ is found, the large $r$ region approaches $AdS_5$ with radius $L$.  This is plausible because $\phi \to 0$ and $\chi \to 0$ for large $r$, and so $V \to -6/L^2$.  The metric is of the form \AdSansatz\ in this region, but with $h(r)$ approaching some positive constant $h_\infty$ which may not be $1$, and with $A(r)$ approaching some linearly increasing function of $r$ which may not be $r/L$.  Suitable coordinate transformations can be made to restore the form of the asymptotic boundary conditions shown in \HybridBCs.  But because our interest is mainly in $T/\Lambda_\phi$ and $\langle {\cal O}_\chi \rangle / \Lambda_\phi^{\Delta_\chi}$, it is convenient to have a way of reading them off from the large $r$ asymptotics without making coordinate transformations.  In fact,
 \eqn{TwoLimits}{\seqalign{\span\TC}{
  \chi(r) \phi(r)^{{\Delta_\chi \over \Delta_\phi - 4}} \to 
   \kappa_\chi = {\langle {\cal O}_\chi \rangle \over 
     \Lambda_\phi^{\Delta_\chi}}  \cr
  {e^{-A(r)} \over \sqrt{h(r)}} \phi(r)^{{1 \over \Delta_\phi-4}}
   {e^{A(r_H)} h'(r_H) \over 4\pi} \to {T \over \Lambda_\phi}
 }}
as $r \to \infty$.  It is easy to check \TwoLimits\ if coordinates are chosen so that the boundary conditions \HybridBCs\ apply.  $T$ is the temperature measured with respect to the time $t$ in \AdSansatz\ with the boundary conditions \HybridBCs\ in force.  Sending $r$ to some linear function of itself preserves the left hand sides of \TwoLimits, and so does $(t,\vec{x}) \to (\lambda t, \lambda \vec{x})$, which is equivalent to adding a constant to $A(r)$.  A similar though slightly more complicated expression could be obtained for $\kappa_\phi = \langle {\cal O}_\phi \rangle / \Lambda_\phi^{\Delta_\phi}$.  But $\langle {\cal O}_\chi \rangle$ is the more interesting quantity because it is the order parameter for spontaneous symmetry breaking.

It would be more interesting to plot $\langle {\cal O}_\chi \rangle/\Lambda_\phi^{\Delta_\chi}$ versus $T/\Lambda_\phi$ than to plot $\chi_0$ versus $\phi_0$, as was done for figure~\ref{figD}.  Unfortunately, in the numerics I have been able to do so far, the quantities in \TwoLimits\ can only be determined at the $10\%$ level due to limited numerical stability.  It would be interesting to push this analysis further because there is a chance of observing non-trivial scaling exponents at the transition.  Perhaps a perturbative treatment of $\chi$ around the critical solution would reveal features of interest.  I hope to report further on these issues in the future.

\subsection{Hair on uncharged black holes, part I}
\label{NEGATIVE}

Is the gauge field an essential part of the construction of hairy black holes given in section~\ref{EXAMPLE}?  Clearly we were able to dispense with it in $AdS_{d+1}$.  There are two further variants of uncharged hairy black holes which deserve mention.  First, if spacetime is asymptotically flat but the potential $V(\phi)$ is negative for large $\phi$, then hairy solutions should exist with the same quantum numbers as a Schwarzschild black hole.  It seems very likely \cite{Hertog:2003ru} that Calabi-Yau compactifications of string theory have a potential which runs away to negative infinity in suitable regions of parameter space, corresponding to putting a positive curvature metric on the internal manifold.  A second way of obtaining hairy uncharged black holes, hinging on a coupling to the square of the Weyl tensor, will be discussed in section~\ref{WEYL}.

\begin{figure}
  \centerline{\includegraphics[width=3in]{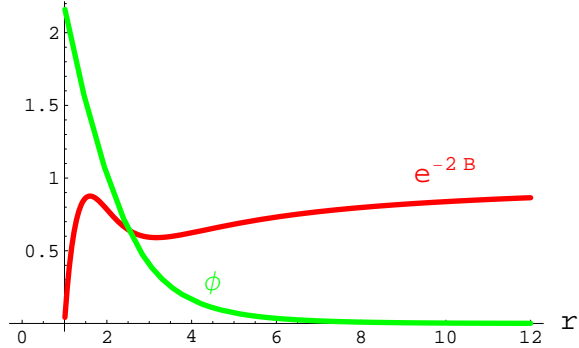}}
  \caption{A hairy black hole that extremizes the action \Vnegative, with $m^2 = 1/5$ and $\lambda = -2/3$.  The horizon is at $r_H = 1$.  The mass of this black hole is $1.61$ times the mass of a Schwarzschild black hole with the same entropy.}\label{figE}
 \end{figure}

The simplest setup to produce uncharged hairy black holes with a potential that is unbounded below is
 \eqn{Vnegative}{\seqalign{\span\TC}{
  S = \int d^4 x \, \sqrt{g} \left[ {R \over 16\pi G_N} - 
   {1 \over 2} (\partial_\mu \phi)^2 - V(\phi) \right]  \cr
  V(\phi) = {m^2 \over 2} \phi^2 + {\lambda \over 4} \phi^4
 }}
with $m^2 > 0$ but $\lambda < 0$.  It is well recognized (see for example \cite{Bekenstein:1996pn}) that to prove scalar no-hair results one generally needs to assume some positive energy condition---for instance, $V(\phi) \geq 0$ everywhere.  So perhaps hairy solutions to the action \Vnegative\ are not particularly novel.  Nevertheless, let us briefly examine some of their features.  As before, units will be chosen so that $M_{\rm Pl} = 1/\sqrt{8\pi G_N} = 1$.  And as before, there is a scaling symmetry, which for the ansatz
 \eqn{QuarticAnsatz}{
  ds^2 = -e^{2A(r)} dt^2 + e^{2B(r)} dr^2 + 
   r^2 (d\theta^2 + \sin^2 \theta d\phi^2) \qquad
  \phi = \phi(r)
 }
can be expressed as
 \eqn{OmegaScaling}{
  t \to \Omega t \quad r \to \Omega r \qquad
  m^2 \to \Omega^{-2} m^2 \qquad 
  \lambda \to  \Omega^{-2} \lambda \,.
 }
It's convenient to use this scaling to set the horizon radius $r_H$ equal to $1$.  The analysis of boundary conditions, both at the horizon and at infinity, proceeds just as in section~\ref{EXAMPLE}.  An example of a hairy black hole is shown in figure~\ref{figE}.  I do not know an example where the hairy black hole has lower mass than the Schwarzschild black hole of the same entropy.

The numerical solution exhibited in figure~\ref{figE} appears to be robust in the sense that changing cutoffs and certain approximation schemes has negligible effect on the solution.  But as a numerical problem, finding hairy black hole solutions that extremize the action \Vnegative\ is a harder than the hairy black hole solutions described in section~\ref{EXAMPLE}.  Heuristically, the reason is that there is barrier penetration involved: $\phi$ must start out at the horizon large enough so that $V(\phi) < 0$ \cite{Gubser:2000nd}, and it must then climb over the barrier of positive $V(\phi)$ to get to $\phi=0$ at infinity.  A typical problem that arises when integrating the differential equations over a fixed range of $r$ is that the scalar appears to be exhibiting the $e^{-mr}/r$ decay but actually spends too long in the $V(\phi) > 0$ region, causing $B'$ eventually to become positive and then singular.  While it is correct in principle to require $\phi \propto e^{-mr}/r$ as $r \to \infty$ as the sole boundary condition at infinity, in practice it is apparently necessary to come up with better finite $r$ criteria.

Because of the difficulties with numerics outlined in the previous paragraph, I am not able to give a description of the phase diagram for this type of hairy black hole.  Having used a choice of dimensions and the scaling symmetry to fix $G_N=1$ and $r_H=1$, there are only two independent parameters, namely $\lambda$ and $m$, and one order parameter, namely $\phi$ at the horizon, which is fixed up to discrete choices once $\lambda$ and $m$ are fixed.  It is reasonable to expect that the transition to hairy black holes is first order: either one tunnels all the way to $V(\phi) < 0$ or keeps $\phi=0$ everywhere.  Holding $\lambda$ fixed and increasing $m$ should eventually prevent hairy solutions from existing, and so it seemed in numerical explorations.  To my suprise, I was unable to find multiple hairy solutions with $\phi>0$ for the same $\lambda$ and $m$---a feature that a phase diagram like figure~\ref{figC} would suggest.

\subsection{Hair on uncharged black holes, part II}
\label{WEYL}

Spacetime has large but finite curvatures near small black hole horizons, so an obvious variant of the previous constructions is to couple the scalar to the curvature tensor.  Couplings of the form $f(\phi) R$ can be removed by a Weyl rescaling at the expense of changing the matter lagrangian.  Thus it makes sense to start with couplings to four-derivative curvature invariants.  There are several of these, but let's focus on the square of the Weyl tensor, $W_{\mu\nu\rho\sigma}^2$.  In imitation of \lagrangian, let's assume an action
 \eqn{WeylSquaredAction}{
  g^{-1/2} {\cal L} &= {R \over 16 \pi G_N} - 
   {1 \over 2} (\partial_\mu \phi)^2 - V(\phi) + 
   {f(\phi) \over 2} W_{\mu\nu\rho\sigma}^2
 }
with $V(\phi) = {1 \over 2} m^2 \phi^2$ and $f(\phi) = \ell^2 \phi^2$ giving the simplest example where non-unique hairy black hole solutions are expected.  A simpler coupling, $\phi W_{\mu\nu\rho\sigma}^2$, does not lead to non-unique black solutions.

\begin{figure}
  \centerline{\includegraphics[width=2.5in]{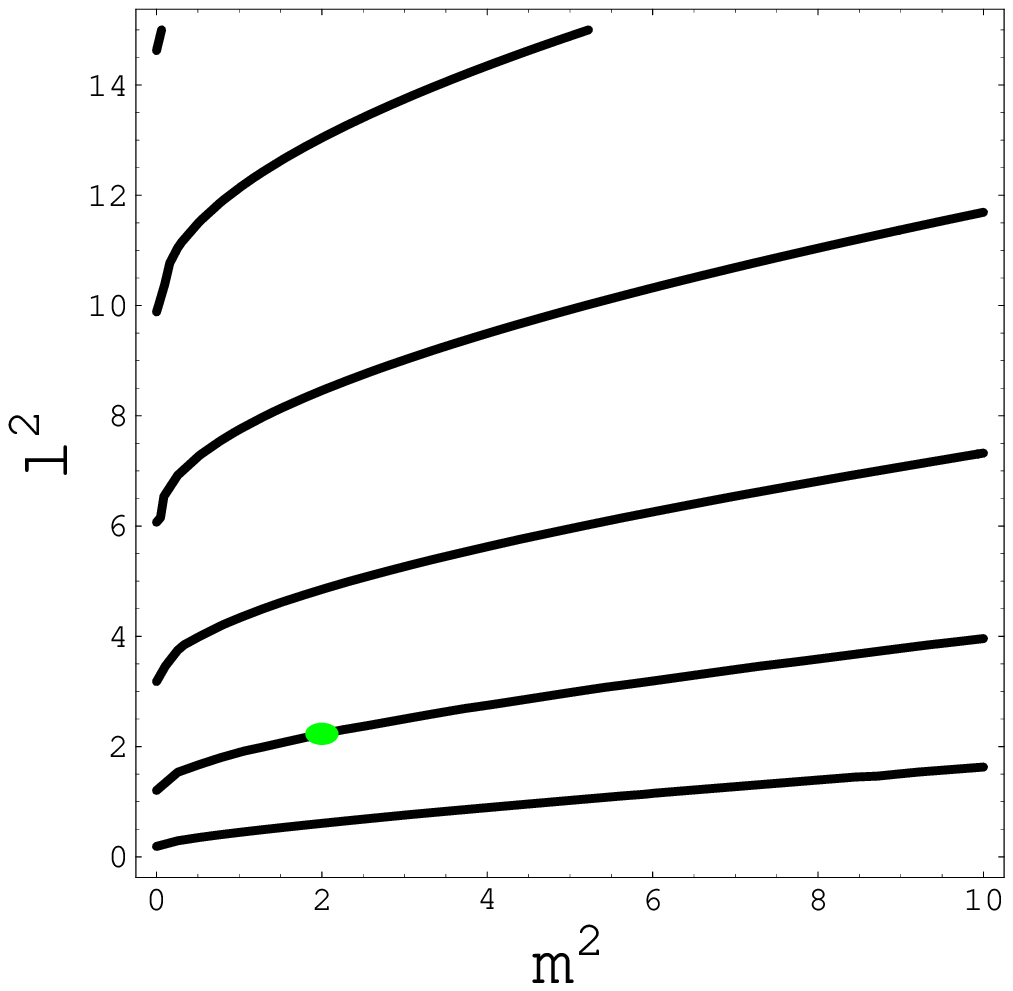}\hskip0.3in
   \raise0.5in\hbox{\includegraphics[width=3in]{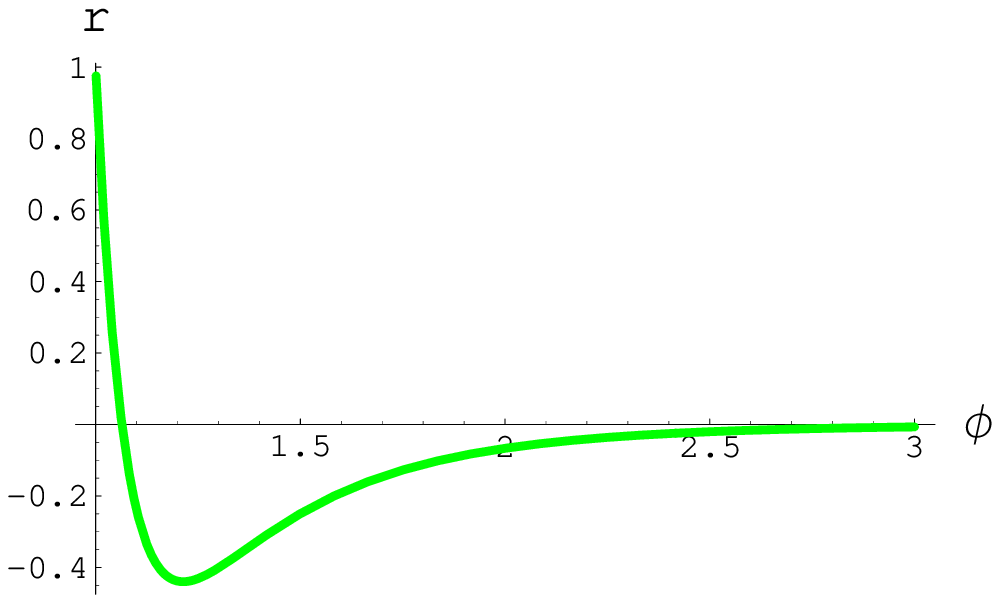}}}\ \\
  \centerline{{\large (A)} \hskip2in {\large (B)}} 
  \caption{(A) The contours show the values of $m^2$ and $\ell^2$ where a normalizable solution to \phiEOM\ occurs.  The lowest contour corresponds to solutions with no nodes.  This contour ends at $m^2=0$ and $\ell^2=0.181$.  Higher contours correspond to increasing numbers of nodes.  (B) The normalizable profile $\phi(r)$ corresponding to the highlighted point on the second contour in (A): $m^2=2$ and $\ell^2=2.231$.}\label{figF}
 \end{figure}

The Schwarzschild solution is still a solution to the theory \WeylSquaredAction.  The simplest way to see that hairy solutions exist is to find normalizable perturbations in $\phi$ around it.  Thus, starting with
 \eqn{dsSchwarzschild}{
  ds^2 = - \left( 1 - {M \over 4\pi r} \right) dt^2 + 
   {dr^2 \over 1 - M/4\pi r} + r^2 d\Omega_2^2
 }
one looks for static solutions of the scalar wave equation,
 \eqn{phiEOM}{
  (\Box - m^2 + \ell^2 W_{\mu\nu\rho\sigma}^2) \phi = 0 \,.
 }
With the usual scaling symmetry arguments in mind, we set $r_H = M/4\pi = 1$ and wind up with two parameters, $m^2$ and $\ell^2$.  For a given value of $m^2$, there is an infinite sequence of values of $\ell^2$ where a normalizable solution occurs.  See figure~\ref{figF}.

\subsection{A flux lattice near the horizon}
\label{LATTICE}

In this section I want to speculate about a highly non-unique black hole construction which is only a slight elaboration of the discussion in section~\ref{GAUGED}.  Consider a theory with a complex scalar whose $U(1)$ phase rotations have been gauged and which has a dimension six coupling to another $U(1)$ gauge field.  The lagrangian is
 \eqn{FLlagrangian}{
  g^{-1/2} {\cal L} &= {R \over 16 \pi G_N} - 
   {1 \over 4e^2} Y_{\mu\nu}^2 - |D_\mu \phi|^2 - 
   {1 \over 4} \left( {1 \over 1 + 2\ell^2 |\phi|^2}
    \right) F_{\mu\nu}^2 - 
   m^2 |\phi|^2  \cr
  Y_{\mu\nu} &= \partial_\mu Y_\nu - \partial_\nu Y_\mu \qquad
   D_\mu \phi = (\partial_\mu + i Y_\mu) \phi
 }
As remarked at the end of section~\ref{GAUGED}, hairy black hole solutions of the type discussed in sections~\ref{EXAMPLE}, \ref{LIMITS}, and~\ref{PHASE} can be embedded in the theory \FLlagrangian\ by fixing the phase of $\phi$ and not turning on any field strength $Y_{\mu\nu}$.  The scalar condensate at the horizon is a superconductor for $Y_{\mu\nu}$.  One may then attempt to generalize the solution further by introducing supercurrents near the horizon: that is, gradients in $\phi$ in the angular directions.  For instance, supercurrents could be set up in the $\phi$ direction on a sphere parameterized so that $ds^2 = d\theta^2 + \sin^2 \theta d\phi^2$.  The result is a magnetic dipole under $Y_{\mu\nu}$ for the black hole.  I do not have a good understanding of whether the supercurrents producing the dipole would eventually run down due to thermal or quantum effects near the horizon.

There is a different situation where the supercurrents can't run down, namely when the black hole carries magnetic monopole charge under $Y_{\mu\nu}$.  Then there must be some number of flux vortices near the horizon to carry away the magnetic flux.  Depending on the values of the parameters, these flux vortices may attract or repel one another.  In a suitable range of parameters with $e$ large, they should repel, as in a type~II superconductor.  Then the minimum energy configurations are roughly described by locating the vortex centers as far from one another as possible.  There are a large number of configurations which are nearly degenerate in energy, differing from one another roughly in how defects are located in an approximately hexagonal lattice.  Evidently, it is difficult to give a fully explicit description of any one solution: all symmetries except time-translation invariance are broken.  However, it might be possible to describe explicitly the simplest cases where there are one or two flux vortices, because then axial symmetry can be retained.

If flux lattices near the horizon indeed exist, then one may inquire what real-time perturbations they have.  It seems to me that energy in time-dependent modes has the opportunity to fall through the horizon, just as argued in section~\ref{CHARGES}.  So if one starts from a non-equilibrium configuration of vortices, the vortices should settle down dissipatively to a configuration which is close to minimal in energy and has no locally unstable perturbations.

A similar construction of flux lattices near horizons could proceed in $AdS_4$.  In that case, for a solution that has the symmetries of the Poincar\'e patch far from the horizon, the flux lattice could be perfectly hexagonal.  Other generalizations might be interesting to explore as well.

\section{Conclusions}
\label{CONCLUSIONS}

It is evident from the examples in this paper that hairy black holes are commonplace in theories with appropriately chosen interactions.  The question is not whether black holes can have hair, but when.

With the exception of section~\ref{NEGATIVE} (and also~\ref{ADSPHASE} when carried over to $AdS_4$) all the examples involved non-renormalizable couplings in the matter lagrangian.  So a natural first guess is that only small black holes can have hair.  This is not quite right because of the scaling symmetry \RigidConformal: it's possible to give scalar hair to black holes of arbitrary size if the mass of the scalar is reduced sufficiently.

An intuitively appealing conjecture which fits the facts is as follows:
 \begin{enumerate}
  \item Perturbatively stable, stationary black hole solutions to a four-dimensional theory whose flat space vacuum state is non-perturbatively stable are uniquely specified by their conserved quantities if the horizon radius is larger than some limit $r_0$.
  \item It is possible to have $r_0 \to \infty$ only if the mass gap $\Delta$ vanishes for bosonic states that can propagate freely in flat space but do not carry a conserved charge of the black hole.
  \item If non-renormalizable terms in the matter lagrangian become significant at a length scale $\ell$, then tuning $\ell \to 0$ also causes $r_0 \to 0$.
  \item A typical situation is to have $r_0 \sim {\ell \over \Delta \sqrt{G_N}}$.
 \end{enumerate}
Points~1 and~3 agree with the findings of \cite{Heusler:1992av,Torii:2001pg} on stable Skyrmion hair.

In short, black hole uniqueness depends on the renormalizability of the matter lagrangian one couples to gravity.  Four is the highest dimension where there are interesting renormalizable theories, and it is the lowest dimension where there are asymptotically flat black holes.  Even without coupling to matter, rotating five-dimensional black holes fail to have the same uniqueness properties as in four dimensions \cite{Emparan:2001wn}.  So the no-hair conjecture is very much tied to four dimensions.

\section*{Acknowledgements}

I thank H.~Verlinde for useful discussions.  This work was supported in part by the Department of Energy under Grant No.\ DE-FG02-91ER40671, and by the Sloan Foundation.

\bibliographystyle{ssg}
\bibliography{hair}

\end{document}